\documentclass[sigconf]{aamas}  

\allowdisplaybreaks
\usepackage{epsfig}
\usepackage{amssymb}
\usepackage{amsfonts}
\usepackage{tikz}
\usepackage{float}
\usepackage{bbm}
\usepackage{lineno}
\usepackage{algorithm}
\usepackage{algpseudocode}
\usepackage{amsmath}

\DeclareMathOperator*{\argmax}{arg\,max}
\newcommand*\circled[1]{\tikz[baseline=(char.base)]{
            \node[shape=circle,draw,inner sep=0.5pt] (char) {#1};}}

\setcopyright{ifaamas}  
\acmDOI{doi}  
\acmISBN{}  
\acmConference[AAMAS'18]{Proc.\@ of the 17th International Conference on Autonomous Agents and Multiagent Systems (AAMAS 2018), M.~Dastani, G.~Sukthankar, E.~Andre, S.~Koenig (eds.)}{July 2018}{Stockholm, Sweden}  
\acmYear{2018}  
\copyrightyear{2018}  
\acmPrice{}  

\begin{abstract}
The spread of unwanted or malicious content through social media has
become a major challenge.
Traditional examples of this include social network spam, but an
important new concern is the propagation of fake news through social
media.
A common approach for mitigating this problem is by using standard statistical
classification to distinguish malicious (e.g., fake news) instances
from benign (e.g., actual news stories).
However, such an approach ignores the fact that malicious instances
propagate through the network, which is consequential both in
quantifying consequences (e.g., fake news diffusing through the
network), and capturing detection redundancy (bad content can be
detected at different nodes).
An additional concern is evasion attacks, whereby the generators of malicious
instances modify the nature of these to escape detection.
We model this problem as a Stackelberg game
between the defender who is choosing parameters of the detection model, and an
attacker, who is choosing both the node at which to initiate malicious
spread, and the nature of malicious entities.
We develop a novel bi-level programming approach for this problem, as
well as a novel solution approach based on implicit function
gradients, and
experimentally demonstrate the advantage of our approach over
alternatives which ignore network structure.
\end{abstract}

\begin{document}
\title{Adversarial Classification on Social Networks}

\author{Sixie Yu}
\affiliation{%
 \institution{Electrical Engineering and Computer Science, Vanderbilt University}
 \city{Nashville}
 \state{TN}
}
\email{sixie.yu@vanderbilt.edu}

\author{Yevgeniy Vorobeychik}
\affiliation{%
  \institution{Electrical Engineering and Computer Science, Vanderbilt University}
 \city{Nashville}
 \state{TN}
}
\email{yevgeniy.vorobeychik@vanderbilt.edu}

\author{Scott Alfeld}
\affiliation{%
 \institution{Computer Science, Amherst College }
 \city{Amherst}
 \state{MA}
}
\email{salfeld@amherst.edu}

\maketitle

\section{Introduction}
\begin{figure}[H]
\centering
\includegraphics[width=2.2in]{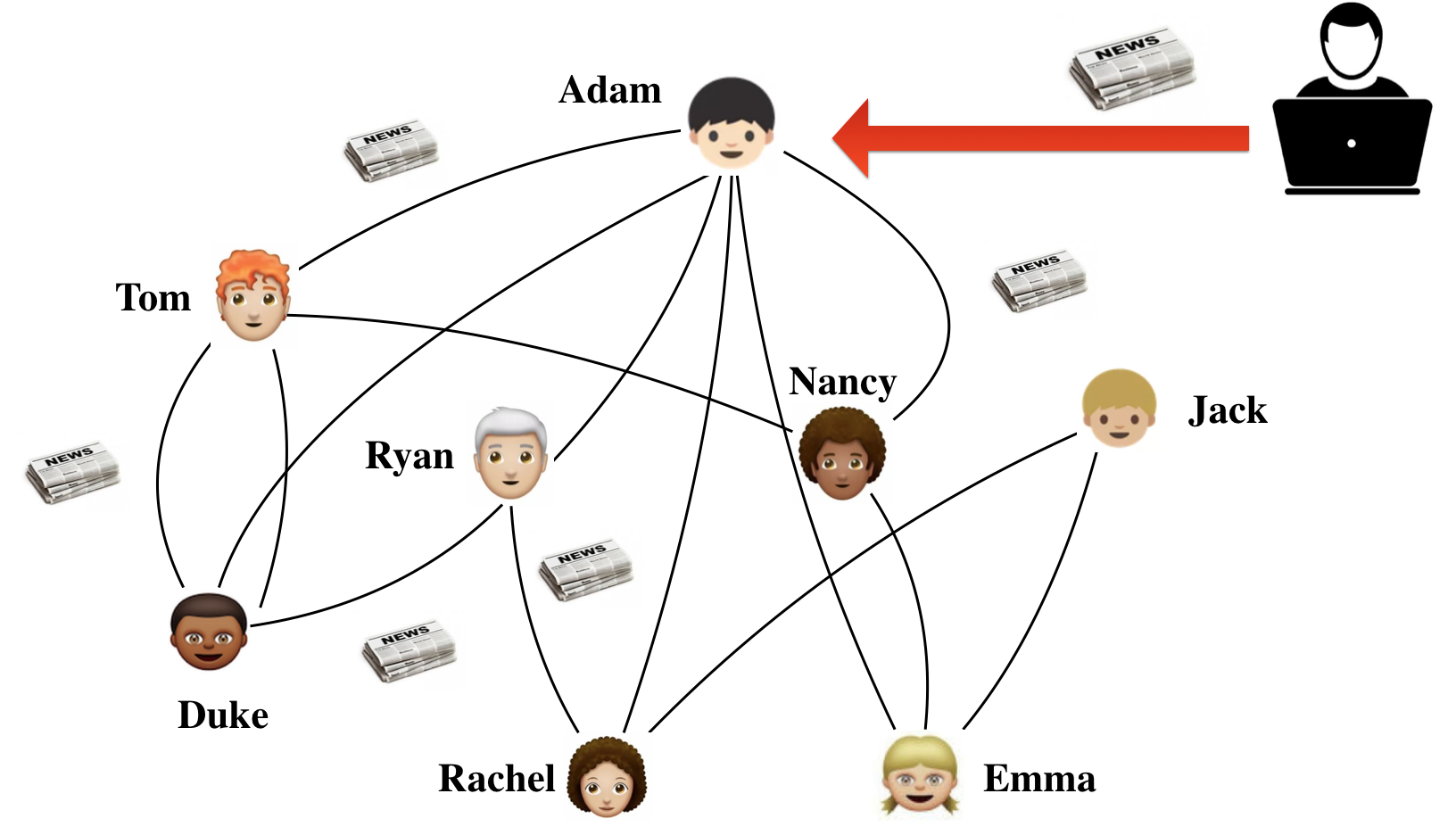}
\caption{An example of the propagation of malicious contents.}
\label{fig:example_attack}
\end{figure}

Consider a large online social network, such as Facebook or Twitter.  It enables unprecedented levels of social interaction in the digital space, as well as sharing of valuable information among individuals.  It is also a treasure trove of potentially vulnerable individuals to exploit for unscrupulous parties who wish to gain an economic, social, or political advantage.
In a perfect world, the social network is an enabler, allowing diffusion of valuable information.
We can think of this ``benign'' information as originating stochastically from some node, and subsequently propagating over the network to its neighbors (e.g., through retweeting a news story), then their neighbors, and so on.
But just as the network is a conduit for valueable information, so it is for ``malicious'' content.
However, such undesirable content can be targeted: first, by selecting an influential starting point on the network (akin to influence maximization), and second, by tuning the content for maximal impact.
For example, an adversary may craft the headline of a fake news story to capture the most attention. 
Consider the illustration in Figure~\ref{fig:example_attack}, where an attacker crafts a fake news story and shares it with Adam. 
This story is then shared by Adam with his friends, and so on.

These are not abstract concerns.
Recently, widespread malicious content (e.g., fake news, antisocial posts) in online social networks has become a major concern. For example, considering that over $50\%$ adults in the U.S. regard social media as their primary sources for news \cite{holcomb2013news}, the negative impact of fake news can be substantial. According to Allcott et al.~\cite{allcott2017social} over 37 million news stories that are later proved fake were shared on Facebook in the last three months of 2016 U.S. presidential election. In addition to fake news, anti-social posts in online communities negatively affect other users and damage community dynamics~\cite{cheng2015antisocial}, while social network spam and phish can defraud users and spread malicious software~\cite{cormack2008email}.

The managers of online social networks are not powerless against these threats, and can deploy detection methods, such as statistical classifiers, to identify and stop the spread of malicious content.
However, such traditional mitigations have not as yet proved adequate.
We focus on two of the reasons for this: first, adversaries can tune content to avoid being detected, and second, traditional learning approaches do not account for network structure.
The implication of network structure mediating both spread and detection has in turn two consequences: first, we have to account for impact of detection errors in terms of benign or malicious content subsequently propgatating through the network, and second, the fact that we can potentially detect malicious content at multiple nodes on the network creates a degree of redundancy.
Consequently, while traditional detection methods use training data to learn a single ``global'' classifier of malicious and benign content, we show that specializing such learning to network structure, and using \emph{different classifiers at different nodes} can dramatically improve performance.

To address the problem of malicious content detection on social networks, we propose two significant modeling innovations.
First, we explicitly model the diffusion process of content over networks \emph{as a function of content} (or, rather, features thereof).
This is a generalization of typical network influence models which abstract away the nature of information being shared.
It is also a crucial generalization in our setting, as it allows us to directly model the balancing act by the attacker between increasing social influence and avoiding detection.
Second, we consider the problem of designing a collection of \emph{heterogeneous} statistical detectors \emph{which explicitly account for network structure and diffusion} at the level of individual nodes, rather than merely training data of past benign and malicious instances.
We formalize the overall problem faced as a Stackelberg game between a defender (manager of the social network) who deploys a collection of heterogeneous detectors, and an attacker who optimally chooses both the starting node for malicious content, and the content itself.
This results in a complex bi-level optimization problem, and we introduce a novel technical approach for solving it, first considering a naive model in which the defender knows the node being attacked, which allows us to develop a projected gradient descent approach for solving this restricted problem, and subsequently utilizing this to devise a heuristic algorithm for tackling the original problem.
We show that our approach offers a dramatic improvement over both traditional homogeneous statistical detection and a common adversarial classification approach.

\paragraph{Related Work}

A number of prior efforts have considered limiting adversarial influence on social networks. 
Most of these pit two influence maximization players against one another, with both choosing a subset of nodes to maximize the spread of their own influence (blocking the influence of the other).
For example, Cerenet  et al. ~\cite{budak2011limiting} consider the problem of blocking a ``bad'' campaign using a ``good'' campaign that spreads and thereby neutralizes the ``bad'' influence.
Similarly, Tsai et al.~\cite{tsai2012security} study a zero-sum game between two parties with competing interests in a networked environment, with each party choosing a subset of nodes for initial influence.
Vorobeychik et al. ~\cite{vorobeychik2015securing} considered an influence blocking game in which the defender chooses from a small set of security configurations for each node, while the attacker chooses an initial set of nodes to start a malicious cascade.
The main differences between this prior work and ours is that (a) our diffusion process depends on the malicious content in addition to network topology, (b) detection at each node is explicitly accomplished using machine learning techniques, rather than an abstract small set of configurations, and (c) we consider an attacker who, in addition to choosing the starting point of a malicious cascade, chooses the content in part to evade the machine learning-based detectors.
The issue of using heterogeneous (personalized) filters was previously studied by Laszka et al.~\cite{Laszka15}, but this work did not consider network structure or adversarial evasion.

Our paper is also related to prior research in single-agent influence maximization and adversarial learning. Kempe et al.~\cite{kempe2003maximizing} initiated the study of influence maximization, where the goal is to select a set of nodes to maximize the total subset of network affected for discrete-time diffusion processes.
Rodriguez et al.~\cite{gomez2012influence} and Du et al.~\cite{du2012learning,du2013uncover,du2013scalable} considered the continuous-time diffusion process to model information diffusion; we extend this model. 
Prior adversarial machine learning work, in turn, focuses on the design of a single detector (classifier) that is robust to evasion attacks~\cite{dalvi2004adversarial, bruckner2011stackelberg,li2014feature}.
However, this work does not consider malicious content spreading over a social network.


\section{Model}


We are interested in protecting a set of agents on a social network from malicious content originating from an external source, while allowing regular (benign) content to diffuse. 
The social network is represented by a graph $G=(V, E)$, where $V$ is the set of vertices (agents) and $E$ is the set of edges.
An edge between a pair of nodes represents communication or influence between them.
For example, an edge from $i$ to $j$ may mean that $j$ can see and repost a video or a news article shared by $i$.
For simplicity, we assume that the network is undirected; generalization is direct.

We suppose that each message (benign or malicious) originates from a node on the network (which may differ for different messages) and then propagates to others. 
We utilize a finite set of instances diffusing over the network (of both malicious and benign content) as a training dataset $D$.  
Each instance, malicious or benign, is represented by a feature vector $x \in \mathbb{R}^n$ where $n$ is the dimension of the feature space. 
The dataset $D$ is partitioned into $D^{+}$ and $D^{-}$, where $D^{+}$ corresponds to malicious and $D^{-}$ to benign instances.


To analyze the diffusion of benign and malicious content on social networks in the presence of an adversary, we develop formal models of (a) the diffusion process, (b) the defender who aims to prevent the spread of malicious content while allowing benign content diffuse, (c) the attacker who attempts to maximize the influence of a malicious message, and (d) the game between the attacker and defender.
We present these next.

\subsection{Continuous-Time Diffusion}
Given an undirected network with a known topology, we use a continuous-time diffusion process to model the propagation of content (malicious or benign) through the social network, extending Rodriguez et al.~\cite{gomez2012influence}. 
In our model, diffusion will depend not merely on the network structure, but also on the nature of the item propagating through the network, which we quantify by a feature vector $x$ as above.

Suppose that the diffusion process for a single message originates at a node $s$.
First, $x$ is transmitted from $s$ to its direct neighbors.
The time taken by a propagation through an edge $e$ is sampled from a distribution over time, $f_{e}(t;\mathbf{w}_e, x)$, which is a function of the edge itself and the entity $x$, and parametrized by $\mathbf{w}_e$.
The affected (influenced) neighbors of $s$ then propagate $x$ to their neighbors, and so on. 
We assume that an affected agent remains affected through the diffusion process.

Given a sample of propagation times over all edges, the time $t_i$ taken to affect an agent $i$ is the length of the shortest path between $s$ and $i$, where the weights of edges are propagation times associated with these edges. 
The continuous-time diffusion process is supplied with a time window  $T$, which is used to simulate time-sensitive natures of propagation, for example, people are generally concerned about a news for several months but not for years. An agent is affected if and only if its shortest path to $s$ is less than or equal to $T$. The diffusion process terminates when the path from $s$ to every unaffected agent is above $T$. 
We define the influence $\sigma(s, x)$ of an instance $x$ initially affecting a network node $s$ as the expected number of affected agents over a fixed time window $T$.

\begin{figure}[h]
\includegraphics[width=2in]{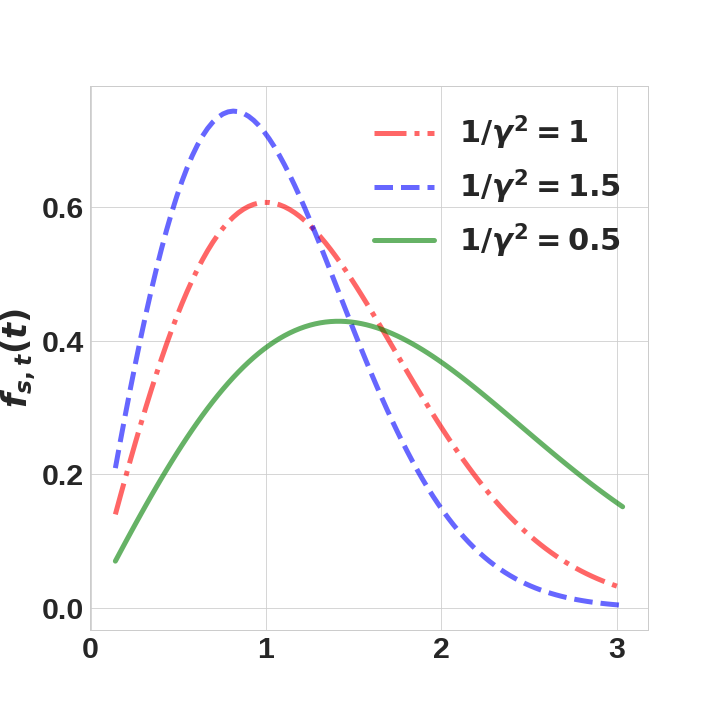}
\caption{Rayleigh distributions with different $1/\gamma^2$.}
\label{fig:rayleigh}
\end{figure}
We assume that the distributions associated with edges are Rayleigh distributions (illustrated in Figure \ref{fig:rayleigh}), which have density function $f(t;\gamma)=\frac{t}{\gamma^2}{e}^{-t^2/(2\gamma^2)}$, where $t\ge 0$ and $\gamma$ is the scale parameter.\footnote{It is straightforward to allow for alternative distributions, such as Weibull.}
The Rayleigh distribution is commonly used in epidemiology and survival analysis \cite{wallinga2004different} and has been recently applied to model information diffusion in social networks \cite{gomez2012influence, du2013uncover}. 
In order to account for heterogeneity among mutual interactions of agents, and to let the influence of a process depend on the content being diffused, we parameterize the Rayleigh distribution of each edge by letting $1/\gamma^2=\mathbf{w}^{T}x$, where $\mathbf{w}$ is sampled from the uniform distribution over $[0,1]$. 
This parameterization results in the following density function for an arbitrary edge:
\begin{equation}\label{eq:param_densityFunc}
f_e(t;\mathbf{w}_e, x)=t({\mathbf{w}_e}^Tx){e}^{-\frac{1}{2}t^2({\mathbf{w}_e}^Tx)}.
\end{equation}
We denote by $\mathcal{W}=\{ \mathbf{w}_e| \forall e \in E \}$ the joint parametrization of all edges. 

Throughout, we assume that the parameters $\mathcal{W}$ are given, and known to both the defender and attacker.
A number of other research studies explore how to learn these parameter vectors from data~\cite{du2012learning, du2013uncover}. 


\subsection{Defender Model}

To protect the network against the spread of malicious content, the network manager---henceforth, \emph{the defender}---can deploy statistical detection, which considers a particular instance (e.g., a tweet with an embedded link) and decides whether or not it is safe.
The traditional way of deploying such a system is to take the dataset $D$ of labeled malicious and benign examples, train a classifier, and use it to detect new malicious content.
However, this approach entirely ignores the fact that a social network mediates the spread of \emph{both} malicious and benign entities.
Moreover, 
both the nature (as captured in the feature vector) and the origin of malicious instances are \emph{deliberate decisions by the adversary} aiming to maximize impact (and harm, from the defender's perspective).
Our key innovations are (a) to allow heterogeneous parametrization of classifiers deployed at different nodes, and (b) to explicitly consider both diffusion and adversarial manipulation during learning.
In combination, this enables us to significantly boost detection effectiveness in social network settings.

Let $\Theta=\{\theta_1, \theta_2, \cdots, \theta_{|V|} \}$ be a vector of parameters of detection models deployed on the network where each $\theta_i \in \Theta$ represents the model used for content shared by node $i$.\footnote{Below, we focus on $\theta_i$ corresponding to detection thresholds as an illustration; generalization is direct.}
We now extend our definition of expected influence  to be a function of detector parameters, denoting it by $\sigma(i, \Theta, x)$, since any content $x$ (malicious or benign) starting at node $i$ which is classified as malicious at a node $j$ (not necessarily the same as $i$) will be blocked from spreading any further.

We define the defender's utility as
\begin{align}\label{eq:U_d}
\begin{split}
U_d = \alpha \sum_{x \in D^{-}}^{} \sum_{i \in V}{\sigma(i, \Theta, x)} - (1-\alpha)\sum_{x \in D^{+}}^{}{\sigma(s, \Theta, z(x))},
\end{split}
\end{align}
where $s$ is the starting node targeted by the adversary, which is subsequently modified by the same adversary into $z(x)$ (in an attempt to bypass detection) when the original content used by the adversary is $x$.
The first part of the utility represents the influence of benign content that the defender wishes to maximize, while the second part denotes the influence of malicious content that the defender aims to limit, with $\alpha$ trading off the relative importance of these two considerations.
Observe that we assume that benign content originates uniformly over the set of nodes, while malicious origin is selected by the adversary.
The defender's action space is the set of all possible parameters $\Theta$ of the detectors deployed at all network nodes.
Note that, as is typical in machine learning, we are using the finite labeled dataset $D$ as a proxy for expected utility with respect to malicious and benign content generated from the same distribution as the data.



\subsection{Attacker Model}


The attacker's decision is twofold: (1) find a node $s \in V$ to start diffusion; and (2) transform malicious content from $x$ (its original, or ideal, form) into another feature vector $z(x)$ with the aim of avoiding detection. 
The first decision is reminiscent of the influence maximization problem\cite{kempe2003maximizing}. 
The second decision is commonly known as the \emph{evasion attack} on classifiers~\cite{lowd2005adversarial, li2014feature}.
In our case, the adversary attempts to balance three considerations: (a) impact, mediated by the diffusion of malicious content, (b) evasion, or avoiding being detected (a critical consideration for impact as well), and (c) a cost of modifying original ``ideal'' content into another form, which corresponds to the associated reduced effectiveness of the transformed content, or effort involved in the transformation.
We impose this last consideration as a hard constraint that $||z(x) - x||_p \le \epsilon$ for an exogenously specified $\epsilon$, where $\|\cdot\|_p$ is the $l_p$ norm.

Consider the collection of detectors with parameters $\Theta$ deployed on the network.
We say that a malicious instance is {\emph detected} at a node $i$ if $\mathbbm{1}[\theta_i(x)=1]=1$, where $\mathbbm{1}(\cdot)$ is the 0-1 indicator function.
The optimization problem of the attacker corresponding to an original malicious instance $x \in D^+$ is then:
\begin{equation}
\begin{aligned}\label{eq:attacker_opt}
& \max_{i, z} & & \sigma(i, \Theta, z)  \\
& s.t & &{||z-x||}_p \le \epsilon  \\
&     & &\mathbbm{1}[\theta_j(z)=1]=0 , \forall j\in V 
\end{aligned}
\end{equation}
where the first constraint is the attacker's budget limit, while the second constraint requires that the attack instance $z$ remains undetected.
If Problem \eqref{eq:attacker_opt} does not have a feasible solution, the attacker sends the original malicious instance without any modification. 
Consequently, the pair $(s,z(x))$ in the defender's utility function above are the solutions to Problem~\eqref{eq:attacker_opt}.

\subsection{Stackelberg Game Formulation}

We formally model the interaction between the defender and the attacker as a Stackelberg game in which the defender is the leader (choosing parameters of node-level detectors) and the attacker the follower (choosing a node to start malicious diffusion, as well as the content thereof).
We assume that the attacker knows $\Theta$, as well as all relevant parameters (such as $\mathcal{W}$) before constructing its attack. 
The equilibrium of this game is the joint choice of $(\Theta, s(\Theta), z(x;\Theta))$, where $s(\Theta)$ and $z(x;\Theta)$ solve Problem~\eqref{eq:attacker_opt}, thereby maximizing the attacker's utility, and $\Theta$ maximizes the defender's utility given $s$ and $z$.
More precisely, we aim to find a Strong Stackelberg Equilibrium (SSE), where the attacker breaks ties in the defender's favor.


We propose finding solutions to this Stackelberg game using the following optimization problem:
\begin{equation}\label{eq:stackelberg_game}
\begin{aligned}
& \max_{\Theta} &  & \alpha \sum_{x \in D^{-}}\sum_i {\sigma(i, \Theta, x)}   - (1-\alpha)\sum_{x \in D^{+}}^{}{\sigma(s, \Theta, z(x))}\\
&s.t.:     &    &     \forall x \in D^+: \quad (s,z(x))\in \argmax_{j, z} \sigma(j, \Theta, z) \\
&     &    &\forall x \in D^+: \quad {||z(x) - x||}_p \le \epsilon  \\
&     &    &\forall x \in D^+: \quad \mathbbm{1}[\theta_k(x)=1]=0 , \forall k\in V
\end{aligned}
\end{equation}
This is a hierarchical optimization problem, where the upper-level optimization corresponds to maximizing the defender's utility.
The constraints of the upper-level optimization are called the lower-level optimization, which is the attacker's optimization problem. 

The optimization problem \eqref{eq:stackelberg_game} is generally intractable for several reasons. 
First, Problem \eqref{eq:stackelberg_game} is a bilevel optimization problem \cite{colson2007overview}, which is hard even when the upper- and lower-level problems are both linear \cite{colson2007overview}. 
The second difficulty lies in maximizing $\sigma(i, \Theta, x)$ (the attacker's problem), as the objective function does not have an explicit closed-form expression.
In what follows, we develop a principled approach to address these technical challenges.

\section{Solution Approach}\label{sec:bilevel-opt}

We start by making a strong assumption that the defender \emph{knows} the node being attacked.
This will allow us to make considerable progress in transforming the problem into a significantly more tractable form.
Subsequently, we relax this assumption, developing an effective heuristic algorithm for computing the SSE of the original problem.

First, we utilize the tree structure of a continuous-time diffusion process to convert (\ref{eq:stackelberg_game}) into a tractable bilevel optimization. We then collapse the bilevel optimization into a single-level optimization problem by leveraging Karush-Kuhn-Tucker (KKT) \cite{boyd2004convex} conditions. 
The assumption that the defender knows the node being attacked allows us to solve the resulting single-level optimization problem using projected gradient descent. 

\subsection{Collapsing the Bilevel Problem}


A continuous-time diffusion process proceeds in a breadth-first-search fashion. It starts from an agent $i$ trying to influence each of its neighbors. Then its neighbors try to influence their neighbors, and so on. 
Notice that once an agent becomes affected, it is no longer affected by others. 
The main consequence of this propagation process is that it results in a propagation tree rooted at $i$, with its structure intimately connected to the starting node $i$.
This is where we leverage the assumption that the defender knows the starting node of the attack: in this case, the tree structure can be pre-computed, and fixed for the optimization problem.

\begin{figure}[h]
\centering
\includegraphics[width=2.4in]{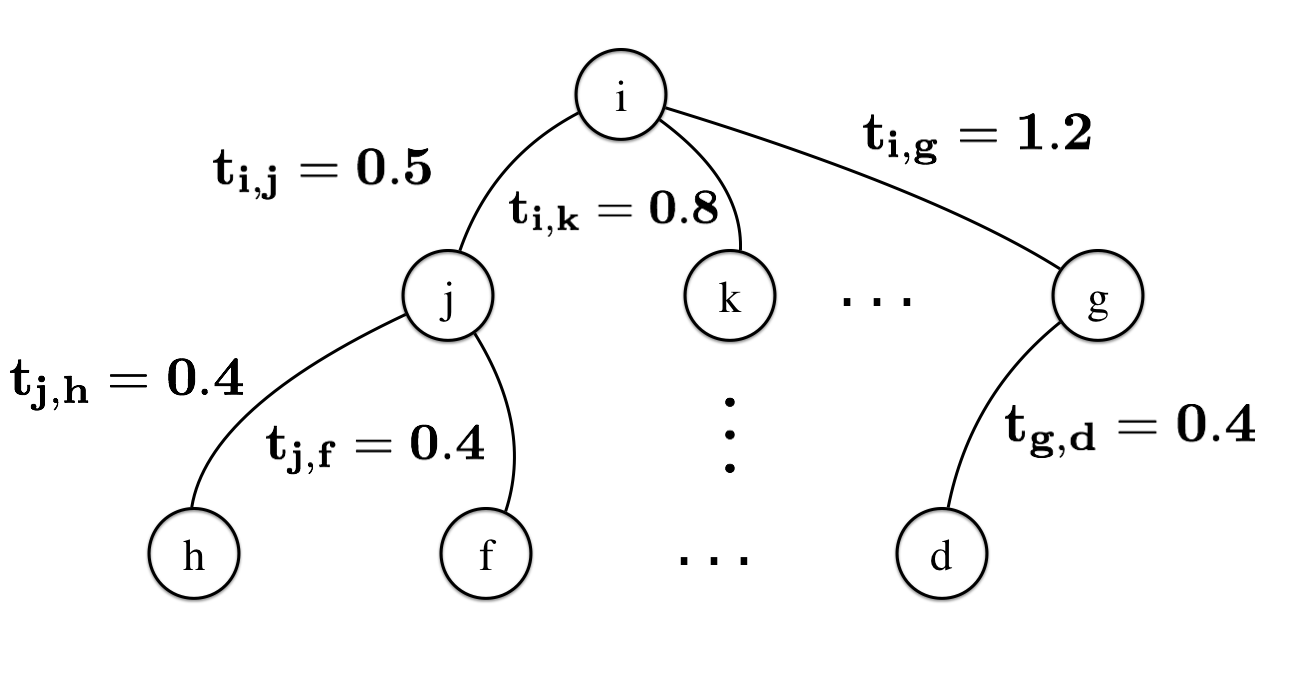}
\caption{A example continuous-time diffusion process.}
\label{fig:diff_process}
\end{figure}

We divide the agents traversed by the tree into several layers in terms of their distances to the source, where each layer is indexed by $l$. 
Since the structure of the tree depends on $i$, $l$ is  a function of $i$, $l(i)$. 
An example of the influence propagation tree is depicted in Figure~\ref{fig:diff_process}, where the first layer consists of $\{ j, k, \cdots, g \}$. 
The number next to each edge represents the weight sampled from the associated distribution. 

We define a matrix ${\mathbf{A}}_l \in \mathbb{R}^{N_l\times n}$  where $N_l$ is the number of agents in layer $l$ and $n$ is the feature dimension of $x$. Each row of $\mathbf{A}_l$ corresponds to the parametrization vector $\mathbf{w}$  of an edge in layer $l$ (an edge is in layer $l$ if one of its endpoint is in layer $l-1$ while the other is in layer $l$; the source is always in layer zero).
For example, in Figure~\ref{fig:diff_process},  $\mathbf{A}_1=[{\mathbf{w}^T_{ij}}; {\mathbf{w}^T_{ik}}; \cdots; {\mathbf{w}^T_{ig}}]$. The product of $\mathbf{A}_l x$ is a vector in $\mathbb{R}^{N_l}$, where each element corresponds to the parameter $1/\gamma^2$ of an edge in layer $l$.

Recall that a sample of random variables from Rayleigh distributions associated with edges corresponds to  a sample of weights associated with these edges. 
With a fixed time window $T$, small edge weights result in wider diffusion of the content over the social network. 
For example, in Figure \ref{fig:diff_process} if the number next to each edge represents a sample of weights, then with $T=1$ the propagation starting from $i$ can only reach agents $j$ and $k$.
However, if  we assume that in another sample  $t_{i,j}, t_{i,k}, t_{i,g}$ all become $0.1$, then with the same time window the propagation can reach every agent in the network.  
Consequently, the attacker's goal is to increase $1/\gamma^2 = \mathbf{w}_e^T x$ for each edge $e$. 
This suggests that in order to increase $1/\gamma^2$ the attacker  can modify the malicious instance $x$ such that the inner products between $x$ and the parameter vectors $\mathbf{w}_e$ of edges are large. 
Consequently, we can formulate the attacker's optimization problem with respect to malicious content $z$ for a given original feature vector $x$ as
\begin{equation}\label{eq:attacker_opt_tractable}
\begin{aligned}
&\max_{z} 	&	&  \sum_{l}^{}{k_l \mathbf{1}^T \mathbf{A}_l z} \\
&s.t.          		    &   &  {||z - x||}_p \le \epsilon \\ 
&						&   &   \mathbbm{1}[\theta_k(z)=1]=0, \forall k\in V.
\end{aligned}
\end{equation}
The attacker aims to make $1/\gamma^2$ for each edge as large as possible, which is captured by the objective function $\mathbf{1}^T \mathbf{A}_l z$, where $\mathbf{1} \in \mathbb{R}^{N_l}$ is a vector with all elements equal to one. Intuitively, this means the attacker is trying to maximize on average the parameter $1/\gamma^2$ of every edge at layer $l$. 
Here, $[k_1, k_2, \cdots, k_l]$ is a vector of decreasing coefficients that provides more flexibility to modeling the attacker's behavior: they are used to re-weight the importance of each layer.
For example, setting $k_1=e^{0}, k_2=e^{-1}, \cdots, k_l=e^{-l}$ models the attacker who tries to make malicious instances spread wider at the earlier layers of the diffusion.

We now use similar ideas to convert the upper-level optimization problem of (\ref{eq:stackelberg_game}) into a more tractable form. 
Suppose that the node being attacked is $s$ (and known to the defender).
Then the defender wants the detection model at $j$ to accurately identify both malicious and benign contents. This is achieved by the two indicator functions inside \circled{1} and \circled{2} in the reformulated objective function of the defender~\eqref{eq:defender_opt_tractable}:
\begin{align}\label{eq:defender_opt_tractable}
\begin{split}
 & \max_{\Theta}{      \underbrace{ \alpha \sum_{x \in D^{-}}^{}\sum_j{ \mathbbm{1}[\theta_j(x)=0]     \sum_{l}^{}{ k_l \mathbf{c}^T_{l,j} \mathbf{A}_l x  }}}_{\circled{1}}     } \\
   & \quad \quad -\underbrace{ (1-\alpha) \sum_{x \in D^{+}}^{}{ \mathbbm{1}[\theta_s(z(x))=0]     \sum_{l}^{}{ k_l \mathbf{c}^T_{l,s} \mathbf{A}_l z(x)  }}}_{\circled{2}}
\end{split}
\end{align}
Notice that this expression includes a vector $\mathbf{c}_{l,j} \in \mathbb{R}^{N_l}$ that does not appear in (\ref{eq:attacker_opt_tractable}).
$\mathbf{c}_{l,j}$ is a function of $\Theta$ and $x$, for a given node $j$ which triggers diffusion (which we omit below for clarity):
\begin{equation}\label{eq:def_c}
\mathbf{c}_{l,j} = \\
\begin{bmatrix}
\mathbbm{1}[\theta_{l_1}(x)=0] \\
\mathbbm{1}[\theta_{l_2}(x)=0] \\
\vdots \\
\mathbbm{1}[\theta_{l_{N_l}}(x)=0]. \\
\end{bmatrix}
\end{equation}
Slightly abusing notation, we let $l_i, i \in [1,2,\cdots,N_l]$ denote the $i$th agent in layer $l$. 
The term $k_l{\mathbf{c}}^T_{l,j} \mathbf{A}_lx$ in \circled{1} can be expanded as follows:
\begin{align}\label{eq:expand}
\begin{split}
k_l  \mathbf{c}^T_{l,j} & \mathbf{A}_lx  \\
& = k_l \begin{bmatrix}
 	\mathbbm{1}[\theta_{l_1}(x)=0], \hdots, \mathbbm{1}[\theta_{l_{N_l}}(x)=0]
     \end{bmatrix}  
     \begin{bmatrix} 
     	\mathbf{w}^T_{l_1} x \\
     	\vdots \\
     	\mathbf{w}^T_{l_{N_l}} x
     \end{bmatrix} \\
     & = k_l \bigg( \mathbbm{1}[\theta_{l_1}(x)=0]\mathbf{w}^T_{l_1}x + \cdots +  \mathbbm{1}[\theta_{l_{N_l}}(x)=0]\mathbf{w}^T_{l_{N_l}}x \bigg),
\end{split}
\end{align}
noting again that $l$ and $N_l$ depend on $j$, the starting node of the diffusion process.
From the expression \eqref{eq:expand}, the defender tries to find $\Theta$ that minimizes the impact of false positives while maximizing the impact of true negatives. 
This is because if each benign instance $x \in D^{-}$ is correctly identified (false-positive rates are zero and true-negative rates are one), the summation at the second line of expression \eqref{eq:expand} will attain its maximum possible value.

In addition to facilitating the propagation of benign contents, the defender wants to limit the propagation of malicious contents, which is embodied in \circled{2}. The equations in \circled{2} are similar to those in \circled{1}, except that the summation is over malicious contents $D^{+}$, and \circled{2} is accounting for the false negatives. 
In this case, $\mathbf{c}_{l,s}$ is a function of $z(x)$, the adversarial feature vector which transforms $x$ into another, $z$. 

We  now  re-formulate the problem  (\ref{eq:stackelberg_game}) as a new bilevel optimization problem (\ref{eq:stackelberg_game_tractable}). 
The upper-level problem corresponds to  the defender's strategy (\ref{eq:defender_opt_tractable}), and the lower-level problem  to the attacker's optimization problem (\ref{eq:attacker_opt_tractable}). 
Here, $s$ is again the node chosen by the attacker.

\begin{align}\label{eq:stackelberg_game_tractable}
\begin{split}
& \min_{\Theta}{ (1-\alpha) \sum_{x \in D^{+}}^{}{ \mathbbm{1}[\theta_s(x)=0]     \sum_{l}^{}{ k_l \mathbf{c}^T_{l,s} \mathbf{A}_l z(x)  }}   }  \\
& \quad \quad  -\alpha \sum_{x \in D^{-}}^{}\sum_j{ \mathbbm{1}[\theta_j(x)=0]     \sum_{l}^{}{ k_l \mathbf{c}^T_{l,j} \mathbf{A}_l x  }} \\
 & \quad \quad \quad \quad s.t: \,\, \forall x \in D^+: z(x) \gets \argmax_{z} \sum_{l}^{}{k_l \mathbf{1}^T \mathbf{A}_l z} \\
 & \quad \quad \quad \quad \quad \quad  s.t. \quad \forall x \in D^+: {||z(x)-x||}_p \le \epsilon \\
 & \quad \quad \quad \quad \quad \quad \quad \quad \,\, \forall x \in D^+:\mathbbm{1}[\theta_k(z(x))=1]=0, \forall k \in V \\
 & \quad \quad \quad \quad \quad \quad \quad \quad \,\, \forall x \in D^+: z(x) \succeq 0,
\end{split}
\end{align}
where the last constraint ensures that $\mathbf{w}^Tz(x) \ge 0$ for all attacks $z(x)$.

The final step, inspired by \cite{mei2015security, mei2015using}, is to convert (\ref{eq:stackelberg_game_tractable}) into a single-level optimization problem via the KKT \cite{boyd2004convex} conditions of the lower-level problem. With appropriate norm constraints (e.g., $l_2$ norm) and a convex relaxation of the indicator functions (i.e., convex surrogates of the indicator functions), the lower-level problem of (\ref{eq:stackelberg_game_tractable}) is convex. 
A convex optimization problem can be equivalently represented by its KKT conditions\cite{burges1998tutorial}.  
The single-level optimization problem then becomes:
\begin{align}\label{eq:single-level-opt}
\begin{split}
& \min_{\Theta}{\quad \hat{F}_d} \\
&  s.t. \quad \forall x:\\
& \partial_{z}\bigg( -\sum_{l}^{}{k_l \mathbf{c}^T_{l,s} \mathbf{A}_l z} + \lambda g(z,x) + \mu^T h(z, \Theta) - \eta^Tz \bigg) = 0 \\
& \qquad  \quad\,\, \lambda g(z, x)= 0, \lambda \ge 0 \\
& \qquad \quad \,\,  g(z,x) \le 0 \\
& \qquad \quad \,\,  \eta \odot (-z) = 0, \eta \succeq 0\\
& \qquad \quad \,\,  h(z, \Theta) = 0 \\
\end{split}
\end{align}
where $\hat{F}_d$ is the objective function of Problem~(\ref{eq:stackelberg_game_tractable}), and $\lambda$, $\mu$, $\eta$ are vectors of lagrangian multipliers. $g(z,x)={||z - x||}_p - \epsilon \le 0$ is the attacker's budget constraint.  
$h(x, \Theta)$ is the set of equality constraints $\mathbbm{1}[\theta_j(z)=1]=0, \forall j \in V$. 
$\eta \odot (-z)$ is the Hadamard (elementwise) product between $\eta$ and $(-z)$ .

\subsection{Projected Gradient Descent}

In this section we demonstrate how to solve the single-level optimization obtained above by projected gradient descent.
The key technical challenge is that we don't have an explicit representation of the gradients with respect to the defender's decision $\Theta$, as these are indirectly related via the optimal solution to the attacker's optimization problem.
We derive these gradients based on the 
implicit function of the defender's utility with respect to $\Theta$.

We begin by outlining the overall iterative projected gradient descent procedure.
In iteration $t$ we update the parameters of detection models by taking a projected gradient step:
\begin{align}\label{eq:project_gradient}
\begin{split}
{\Theta}^{(t+1)} = \text{Proj}_{\mathcal{A}_d}\bigg( \Theta^{(t)} - \beta_t {\nabla}_{\Theta}\hat{F}_d\big|_{\Theta=\Theta^{(t)}}  \bigg)
\end{split}
\end{align}
where $\mathcal{A}_d$ is the feasible domain of $\Theta$ and $\beta_t$ is the learning rate. 
With ${\Theta}^{(t+1)}$ we solve for $z^{(t+1)}$, which is the optimal attack for a fixed ${\Theta}^{(t+1)}$.
${\nabla}_{\Theta}\hat{F}_d$ is the gradient of the upper-level problem.

Expanding ${\nabla}_{\Theta}\hat{F}_d$ using the chain rule and still using $s$ as the initially attacked node, we obtain
\begin{align}
\begin{split}\label{eq:dF_dTheta}
& {\nabla}_{\Theta}\hat{F}_d   =  (1-\alpha)\circled{1} - \alpha\circled{2} \\
\circled{1}  = & \sum_{ x \in D^{+} }^{}{  \bigg[ \frac{\partial \mathbbm{1}[\theta_s(z(x))=0]}{\partial \Theta}\sum_{l}^{}{k_l \mathbf{c}^T_{l,s} \mathbf{A}_l z(x)}  } + \\
& \qquad  \mathbbm{1}[\theta_s(z(x))=0] \underbrace{\frac{\partial [\sum_{l}^{}{k_l \mathbf{c}^T_{l,s} \mathbf{A}_l z(x)}]}{\partial \Theta}}_{\text{(a)}} \bigg]  \\
\circled{2}  = & \sum_{x \in D^{-}}^{}\sum_j{ \bigg[ \frac{\partial \mathbbm{1}[\theta_j(x)=0]}{\partial \Theta}\sum_{l}^{}{k_l \mathbf{c}^T_{l,j} \mathbf{A}_l x}} + \\
& \qquad \mathbbm{1}[\theta_j(x)=0] \underbrace{\frac{\partial [\sum_{l}^{}{k_l \mathbf{c}^T_{l,j} \mathbf{A}_l x}]}{\partial \Theta}}_{\text{(b)}} \bigg]  
\end{split}
\end{align}
In both \circled{1} and \circled{2} we note that $\frac{\partial \mathbbm{1}[\theta_j(x)=0]}{\partial \Theta}$ is dependent on the specific detection models. We will give a concrete example of their derivation in Section \ref{sec:logistic}. 

In $\sum_{l}^{}{k_l \mathbf{c}^T_{l,s} \mathbf{A}_l z(x)}$ there are two terms that are functions of $\Theta$: $\mathbf{c}_{l,s}$ and $z(x)$. Consequently, $(a)$ can be expanded as:
\begin{align}\label{eq:(a)}
\begin{split}
(a) = \sum_{l}^{} {k_l \bigg[   \frac{\partial \mathbf{c}_{l,s}} {\partial \Theta_l}\mathbf{A}_l z(x) + {\left[\frac{\partial z(x)}{\partial \Theta_l}\right]}^T \mathbf{A}^T_l \mathbf{c}_{l,s} \bigg]}.
\end{split}
\end{align}
Note that only the detection models of those agents at layer $l$ have contribution to $\mathbf{c}_{l,s}$. Thus, $\frac{\partial \mathbf{c}_{l,s}}{\partial \Theta_l}$ is a Jacobian matrix with dimension $N_l \times N_l$, where $N_l$ is the number of agents at layer $l$ and $\Theta_l$ denotes the detection models of those $N_l$ agents. Since $\mathbf{c}_{l,s}$ is also dependent on the specific detection models of  agents, we defer its derivation to Section \ref{sec:logistic}. 

$\frac{\partial z(x)}{\partial \Theta_l}$ is a $n\times N_l$ Jacobian matrix and is the main difficulty  because we do not have an explicit function of the attacker's optimal decision $z(x)$ with respect to $\Theta_l$. 
Fortunately, the constraints in (\ref{eq:single-level-opt}) implicitly define $z(x)$ in terms of $\Theta$:
\begin{align}
\begin{split}
& \mathbf{f}(\Theta, z, \lambda, \mu, \eta) =  \\
& \begin{bmatrix}
\partial_{z} \bigg( -\sum_{l}^{}{k_l \mathbf{c}^T_{l,s} \mathbf{A}_l z} + \lambda g(z,x) + \mu^T h(z, \Theta) - \eta^Tz \bigg) \\
\lambda g(z,x) \\
\mu^T h(z, \Theta) \\
\eta \odot (-z)
\end{bmatrix}
\end{split}
\end{align}
$\Theta$ and the attacked malicious instance $z$ satisfy $\mathbf{f}(\Theta, z, \lambda, \mu, \eta) = \mathbf{0}$. 
The Implicit Function Theorem\cite{zorichmathematical} states that if $\mathbf{f}(\Theta, z, \lambda, \mu, \eta)$ is continuous and differentiable and the Jacobian matrix 
$$\left[ \frac{\partial \mathbf{f}}{\partial z}|  \frac{\partial \mathbf{f}}{\partial \lambda}| \frac{\partial \mathbf{f}}{\partial \mu}| \frac{\partial \mathbf{f}}{\partial \eta}\right]$$
has full rank, there is a unique implicit function $I(\Theta)=(z, \lambda, \mu, \eta)$.
Moreover, the derivative of $\frac{\partial I}{\partial \Theta}$ is:
\begin{align}
\begin{split}\label{eq:dI_dTheta}
\frac{\partial I}{\partial \Theta} = -{\begin{bmatrix}
\frac{\partial \mathbf{f}}{\partial z} | \frac{\partial \mathbf{f}}{\partial \lambda} | \frac{\partial \mathbf{f}}{\partial \mu} | \frac{\partial \mathbf{f}}{\partial \eta} 
\end{bmatrix}}^{-1} \left( \frac{\partial \mathbf{f}}{\partial \Theta} \right).
\end{split}
\end{align}
$\frac{\partial \mathbf{f}}{\partial z}$ is the Jacobian matrix of $\mathbf{f}(\Theta, z, \lambda, \mu, \eta)$ with respect to $z$, and so on. $\frac{\partial z}{\partial \Theta} \in \mathbbm{R}^{n\times N}$ is the first $n$ rows of $\frac{\partial I}{\partial \Theta}$, where $\frac{\partial z}{\partial \Theta_l}$ can be column-wise indexed by the nodes at layer $l$. 

$(b)$ can be similarly expanded as we had done for $(a)$, except that the attacker does not modify benign content, so that $x \in D^{-}$ is no longer a function of $\Theta$:
\begin{align}\label{eq:(b)}
\begin{split}
(b) = \sum_{l}^{}\sum_j{k_l \bigg[   \frac{\partial \mathbf{c}_{l,j}}{\partial \Theta_l}\mathbf{A}_l x \bigg]}.
\end{split}
\end{align}
The full projected gradient descent approach is given by Algorithm~\ref{algo:find_defense}.
\begin{algorithm}
\caption{Find Defense Strategy}\label{algo:find_defense}
\begin{algorithmic}[1]
\State \textbf{Input}: agent $j$
\State \textbf{Initialize}: $\Theta^{(0)}, \lambda, \mu, \eta, \beta_0$
\For{$t=1\cdots k$}  
	\State $\Theta^{(t+1)}=\text{Proj}_{\mathcal{A}_d}\bigg( \Theta^{(t)} - \beta_t {\nabla}_{\Theta}\hat{F}_d\big|_{\Theta=\Theta^{(t)}}  \bigg)$ 
\EndFor \\
\Return $\Theta^{(k+1)}$
\end{algorithmic}
\end{algorithm}

\subsection{Optimal Attack}

So far, we had assumed that the network node being attacked is fixed.
However, the ultimate goal is to allow the attacker to choose both the node $s$, and the modification of the malicious content $z$.
We begin our generalization by first allowing the attacker to optimize these jointly.

The full attacker algorithm which results is described in Algorithm \ref{algo:opt_attack}.
\begin{algorithm}[H]
\caption{Optimal Attack Strategy}\label{algo:opt_attack}
\begin{algorithmic}[1]
\State \textbf{Input}: $\Theta, x$
\State \textbf{Initialize}: $ret=[]$
\For{$i=1\cdots |V|$}  
	\State $x(i) \gets \text{Solve (\ref{eq:attacker_opt_tractable})}$ 
	\State $\hat{U}_a(i) \gets \text{Optimal objective value of (\ref{eq:attacker_opt_tractable})}$
	\State $\big(i, z(i,x),\hat{U}_a(i) \big)$ appended to $ret$
\EndFor
\State $z, s \gets \text{OptimalTuple(ret)}$ \\
\Return $z, s$
\end{algorithmic}
\end{algorithm}
Recall that the tree structure of a propagation is dependent on the agent being attacked, which makes the objective function of (\ref{eq:attacker_opt_tractable}) a function of the agent being attacked. 
Thus, for a given fixed $\Theta$, the attacker iterates through each agent $i$ and solves the problem~\eqref{eq:attacker_opt_tractable}, assuming the propagation starts from $i$, resulting in associated utility $\hat{U}_a(i)$ and an attacked instance $z(i,x)$. Then $i$, $z(i,x)$, and $\hat{U}_a(i)$ are appended into a list of a 3-tuples (the sixth step in Algorithm \ref{algo:opt_attack}). 
When the iteration completes the attacker picks the optimal 3-tuple in terms of utility (eighth step in Algorithm \ref{algo:opt_attack}, where the function \textit{OptimalTuple(ret)} finds the optimal 3-tuple from the list \textit{ret}). 
The node $s$ and the corresponding attack instance $z$ in this optimal 3-tuple become the optimal attack.

\subsection{SSE Heuristic Algorithm}

Now we take the final step, relaxing the assumption that the attacker chooses a fixed node to attack which is known to the defender prior to choosing $\Theta$.
Our main observation is that fixing $s$ in the defender's algorithm above allows us to find a collection of heterogeneous detector parameters $\Theta$, and we can evaluate the \emph{actual} utility of the associated defense (i.e., if the attacker optimally chooses both $s$ and $z$ in response) by using Algorithm~\ref{algo:opt_attack}.
We use this insight to devise a simple heuristic: iterate over all potential nodes $s$ that can be attacked, compute the associated defense $\Theta(s)$ (using the optimistic definition of defender's utility in which $s$ is assumed fixed), then find the actual optimal attack in response for each $x \in D^+$.
Finally, choose the $\Theta(s)$ which has the best \emph{actual} defender utility.

This heuristic algorithm is described in Algorithm \ref{algo:opt_defense}.
\begin{algorithm}
\caption{Optimal Defense Strategy}\label{algo:opt_defense}
\begin{algorithmic}[1]
\State \textbf{Input}: $G=(V,E), \mathcal{W}, D$
\For{$j=1\cdots |V|$}  
	\State $\Theta_j \gets \text{Apply Algorithm \ref{algo:find_defense}}$
	\State $\forall x \in D^+: (s,z(x)) \gets \text{Apply Algorithm \ref{algo:opt_attack}}$ 
	\State $\hat{U}_d(j) \gets \textit{DefenderUtility}(\Theta_j, (s,z(x)))$
\EndFor
\State $j \gets \argmax_{j}{\hat{U}_d(j)}$ \\
\Return $\Theta_j$
\end{algorithmic}
\end{algorithm}
The fifth step in the algorithm includes the function \textit{DefenderUtility}, which evaluates the defender's utility $\hat{U}_d(j)$. 
Note that the input argument $s$ of this function is used to determine the tree structure of the propagation started from $s$. 

Recall that Algorithm \ref{algo:find_defense} solves (\ref{eq:single-level-opt}), which depends on the specific detection model to compute the relevant gradients. 
Therefore, in what follows, we present a concrete example of how to solve (\ref{eq:single-level-opt}) where detection models are logistic regressions. Specifically, we illustrate how to derive the two terms, $\frac{\partial \mathbbm{1}[\theta_j(z)=0]}{\partial \Theta}$ and $\frac{\partial \mathbf{c}_{l,j}}{\partial \Theta_l}$ that depend on particular details of the detection model.

\subsection{Illustration: Logistic Regression}\label{sec:logistic}

We consider the logistic regression model used for detection at individual nodes to illustrate the ideas developed above.
For a node $i$, its detection model has two components: the logistic regression $\frac{1}{1 + e^{-\phi^T x}}$, where $\phi$ is the weight vector of the logistic regression and $x$ the instance propagated to $i$, and a detection threshold $\theta_i$ (which is the parameter the defender will optimize).
An instance is classified as benign if 
$\frac{1}{1 + e^{-\phi^T x}} \le \theta_i$.
Thus (slightly abusing notation as before), $\theta_i(x) \ne 0$ ($x$ is classified as malicious) if $\frac{1}{1 + e^{-\phi^T x}} \ge \theta_i$.

With the specific forms of the detection models we can derive $\frac{\partial \mathbbm{1}[\theta_j(x)=0]}{\partial \Theta}$ and $\frac{\partial \mathbf{c}_l}{\partial \Theta_l}$ (omitting the node index $s$ or $j$ for clarity).  
A technical challenge is that the indicator function $\mathbbm{1}(\cdot)$ is not continuous or differentiable, which means that it's difficult to characterize its derivative with respect to $\Theta$. 
However, observe that for logistic regression $\theta_j(x) = 0$ $\big( \frac{1}{1 + e^{-{\phi}^T x}}  \le \theta_j\big)$   is equivalent to $\log\big(  \frac{\theta_j}{1-\theta_j} \big) \ge \phi^Tx$. 
Therefore we use $\log\big(  \frac{\theta_j}{1-\theta_j} \big) - \phi^Tx$ as a surrogate function for $\mathbbm{1}[\cdot]$. 
Then $\frac{\partial \mathbbm{1}[\theta_j(x)=0]}{\partial \Theta}$ is a $N$-dimension vector with the $j$th element equal to $\frac{1}{\theta_j - \theta^2_j}$. The $\mathbf{c}_l$ vector then becomes:
\begin{equation}\label{eq:def_c_updated}
\mathbf{c}_l = \\
\begin{bmatrix}
 \log\big(  \frac{\theta_{l_1}}{1-\theta_{l_1}} \big) - \phi^Tx \\
 \log\big(  \frac{\theta_{l_2}}{1-\theta_{l_2}} \big) - \phi^Tx \\
\vdots 
\end{bmatrix}
\end{equation}
and $\frac{\partial \mathbf{c}_l}{\partial \Theta_l}$ becomes a $N_l \times N_l$ diagonal matrix:
\begin{equation}\label{eq:dCl_dTheta}
\frac{\partial \mathbf{c}_l}{\partial \Theta_l} = \\
		  \begin{bmatrix}
		      \frac{1}{{\theta}_{l_1}-{\theta}^2_{l_1}}  & & \\
		    & \ddots & \\
		    & &   \frac{1}{{\theta}_{N_l}-{\theta}^2_{N_l}} 
		  \end{bmatrix}
\end{equation}
With equations (\ref{eq:dF_dTheta})-(\ref{eq:(b)}), $\frac{\partial \mathbf{c}_l}{\partial \Theta_l}$ and $\frac{\partial \mathbbm{1}[\theta_j(x)=0]}{\partial \Theta}$, we can now calculate $\nabla_{\Theta}\hat{F}_d$. 
Since the thresholds $\theta_i \in [0, 1]$, the defender's action space is $[0,1]^N$. 
When updating $\Theta$ by (\ref{eq:project_gradient}) we therefore project it back to $[0,1]^N$ in each iteration. 

\section{Experiments}\label{sec:exp}
In this section we experimentally evaluate our proposed approach.
We used the Spam dataset \cite{Lichman:2013} from UCI machine learning repository as the training dataset for the logistic regression model. 
The Spam dataset contains 4601 emails, where each email is represented by a 57-dimension feature vector. We divided the dataset into three disjoint subsets: $D'$ used to learn the logistic regression (tuning the weight vectors with thresholds setting to $0.5$) as well as other models to which we compare, $D_{\text{train}}$ used in Algorithm \ref{algo:opt_defense} to find the optimal defense strategy, and $D_{\text{test}}$ to test the performance of the defense strategy. 
The sizes of $D'$, $D_{\text{train}}$, and $D_{\text{test}}$ are 3681, 460, and 460, respectively. They are all randomly sampled from $D$.  

Our experiments were conducted on two synthetic networks with 64 nodes: Barabasi-Albert preferential attachment networks (BA)~\cite{barabasi1999emergence} and Watts-Strogatz networks (Small-World) ~\cite{watts1998collective}. BA is characterized by its power-law degree distribution, where connectivity is heavily skewed towards high-degree nodes. 
The power-law degree distribution, $P(k)\sim k^{-r}$, gives the probability that a randomly selected node has $k$ neighbors. 
The degree distributions of many real-world social networks have previously been shown to be reasonably approximated by the power-law distribution with $r \in [2.1,2.4]$ \cite{barabasi2002evolution}. 
Our experiments for BA were conducted across two sets of parameters: $r=2.1$ and $r=2.3$. 

The Small-World topology is well-known for balancing shortest path distance between pairs of nodes and local clustering in a way as to qualitatively resemble real networks~\cite{ugander2011anatomy}.
In our experiments we consider two kinds of Small-World networks.
The first has average length of shortest path equal to $5.9$ and local clustering coefficient equal to $0.144$. In this case the local clustering coefficient is close to what had been observed in large-scale Facebook friendship networks~\cite{ugander2011anatomy}. 
The second one has average shortest path length of $5$ and local clustering coefficient of $0.08$, where the local clustering coefficient is close to that for the electric power grid of the western United States \cite{watts1998collective}. 

Our node-level detectors use logistic regression, with our algorithm producing the threshold for these.
The trade-off parameter $\alpha$ was set to $0.5$ and the time window $T$ was set to $1$. We applied standard pre-processing techniques to transform each feature to lie between zero and one. The attacker's budget is measured by squared $l_2$ norm and the budget limit $\epsilon$ is varied from $0.001$ to $0.01$. 
We compare our strategy with three others based on traditional approaches: \emph{Baseline}, \emph{Re-training}, and \emph{Personalized-single-threshold}; we describe these next. 

\textbf{Baseline}: This is the typical approach which simply learns a logistic regression on training data, sets all thresholds to $0.5$, and deploys this model at all nodes.

\textbf{Re-training}: The idea of re-training, common in adversarial classification, is to iteratively augment the original training data with attacked instances, re-training the logistic regression each time, until convergence~\cite{barreno2006can, li2016general}. 
The logistic regressions deployed at the nodes are homogeneous, with all thresholds being $0.5$. 

\textbf{Personalized-single-threshold}: This strategy is only allowed to tune a single agent's threshold. It has access to  $D_{\text{train}}$ that includes unattacked emails. The strategy iterates throught each node $i$ and finds its optimal threshold. The optimality is measured by the defender's utility as defined in (\ref{eq:U_d}), where the expected influence of an instance  is estimated by simulating 1000 propagations started from $i$. Then the strategy picks the node with largest utility and sets its optimal threshold.

As stated earlier, network topologies and  parameter vectors associated with edges are assumed to be known by both the defender and the attacker. 
The attacker has full knowledge about the defense strategy, including the weight vectors of logistic regressions as well as their thresholds. As in the definition of Stackelberg game, the evaluation procedure lets the defender first choose its strategy $\Theta^{\ast}$, and then the attacker computes its best response, which chooses the initial node for the attack $s$ and transformations of malicious content $z$ aimed at evading the classifier.
Finally  the defender's utility is calculated by (\ref{eq:U_d}),
where the expected influence is estimated by simulating 1000 propagations originating from $s$ for each malicious instance $z$. 

\begin{figure}[h]
\begin{tabular}{cc}
\includegraphics[width=1.6in]{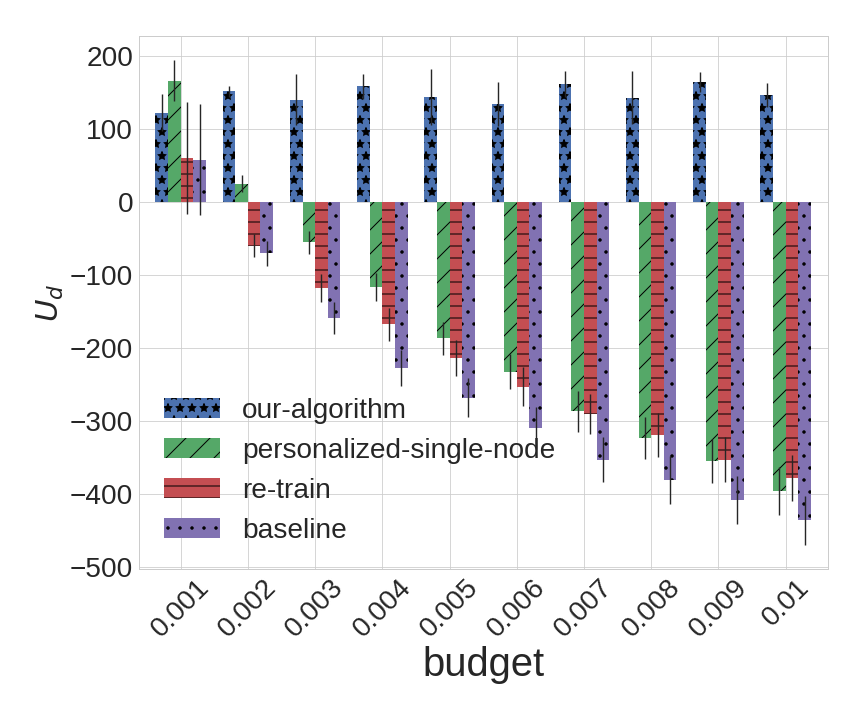} &
\includegraphics[width=1.6in]{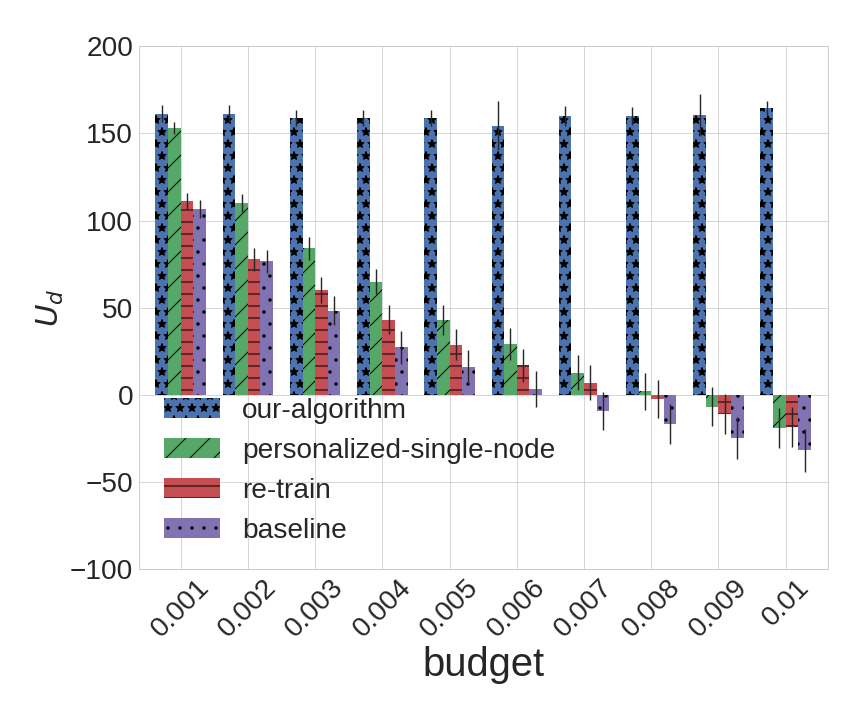}
\end{tabular}
\caption{The performance of each defense strategy. Each bar is averaged over 10 random topologies. Left: BA. Right: Small-world) }
\label{fig:exp_ret_1}
\end{figure}

\begin{figure}[h]
\begin{tabular}{cc}
\includegraphics[width=1.6in]{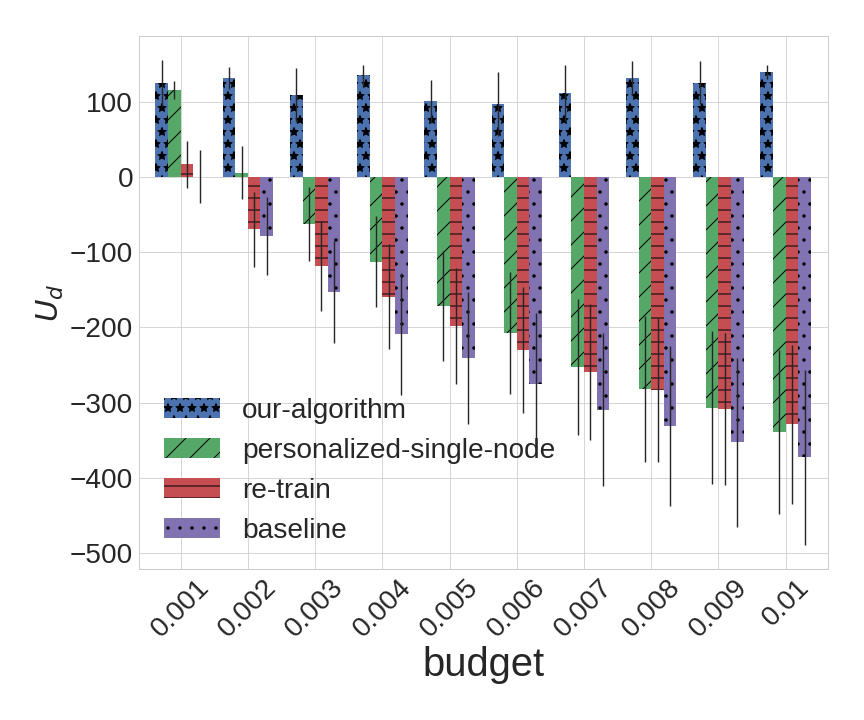} &
\includegraphics[width=1.6in]{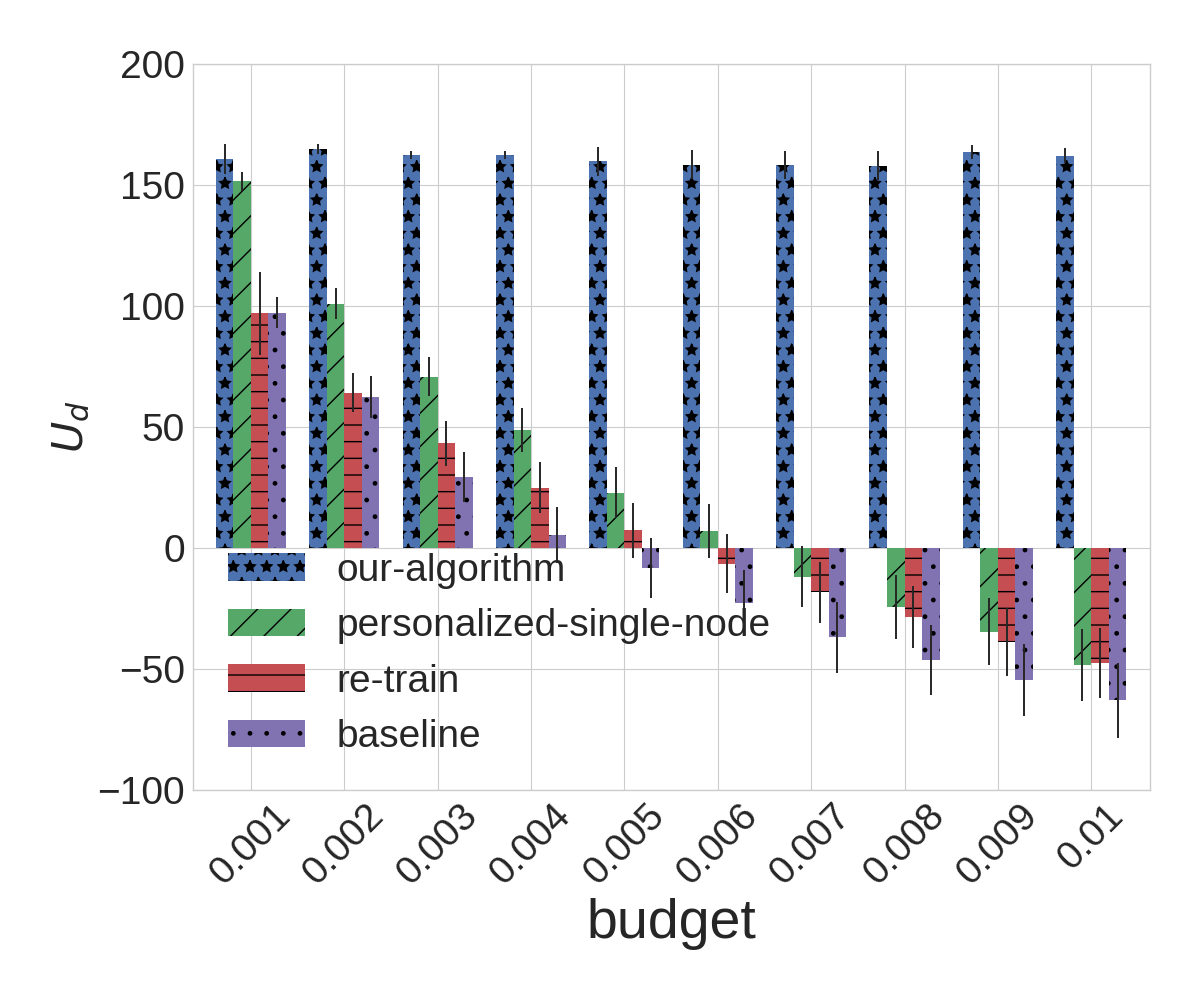}
\end{tabular}
\caption{The performance of each defense strategy. Each bar is averaged over 10 random topologies. Left: BA. Right: Small-world) }
\label{fig:exp_ret_2}
\end{figure}

The experimental results for BA ($r=2.1$) and Small-World (average length of shortest path=$5.9$ and local clustering coefficient=$0.144$) are shown in Figure \ref{fig:exp_ret_1}, and the experimental results for BA ($r=2.3$) and Small-World (average length of shortest path=$5$ and local clustering coefficient=$0.08$) are shown in Figure \ref{fig:exp_ret_2}.

As we can observe from the experiments, our algorithm outperforms all of the alternatives in nearly every instance; the sole exception is when the attacker budget is $0.001$, which effectively eliminates the adversarial component from learning.
For larger budgets, our algorithm remarkably robust even as other algorithms perform quite poorly, so that when $\epsilon = 0.01$, there is a rather dramatic gap between our approach and all alternatives.
Not surprisingly, the most dramatic differences can be observed in the BA topology: with the large variance in the degree distribution of different nodes, our heterogeneous detection is particularly valuable in this setting.
In contrast, the degradation of the other methods on Small-World topologies is not quite as dramatic, although the improvement offered by the proposed approach is still quite pronounced.
Among the alternatives, it is also revealing that personalizing thresholds results in second-best performance: again, takng account of network topology is crucial; somewhat surprisingly, it often outperforms re-training, which explicitly accounts for adversarial evasion, but not network topology.


\section{Conclusion}

We address the problem of adversarial detection of malicious content spreading through social networks.
Traditional approaches use with a homogeneous detector or a personalized filtering approach. Both ignore (and thus fail to exploit knowledge of) the network topology, and most filtering approaches in prior literature ignore the presence of an adversary.
We present a combination of modeling and algorithmic advances to systematically address this problem.
On the modeling side, we extend diffusion modeling to allow for dependence on the \emph{content} propagating through the network, model the attacker as choosing both the malicious content, and initial target on the social network, and allow the defender to choose heterogeneous detectors over the network to block malicious content while allowing benign diffusion.
On the algorithmic side, we solve the resulting Stackelberg game by first representing it as a bilevel program, then collapsing this program into a single-level program by exploiting the problem structure and applying KKT conditions, and finally deriving a projected gradient descent algorithm using explicit and implicit gradient information.
Our experiments show that our approach dramatically outperforms, homogeneous classification, adversarial learning, and heterogeneous but non-adversarial alternatives.

\section{Acknowledgements}
This work was supported, in part by the National Science Foundation (CNS-1640624, IIS-1649972, and IIS-1526860), Office of Naval Research (N00014-15-1-2621) and Army Research Office (W911NF-16-1-0069).

%
%
\bibliographystyle{abbrv}
\bibliography{ms.bib}

\begin{thebibliography}{10}

\bibitem{allcott2017social}
H.~Allcott and M.~Gentzkow.
\newblock Social media and fake news in the 2016 election.
\newblock Technical report, National Bureau of Economic Research, 2017.

\bibitem{barabasi1999emergence}
A.-L. Barab{\'a}si and R.~Albert.
\newblock Emergence of scaling in random networks.
\newblock {\em Science}, 286(5439):509--512, 1999.

\bibitem{barabasi2002evolution}
A.-L. Barab{\^a}si, H.~Jeong, Z.~N{\'e}da, E.~Ravasz, A.~Schubert, and
  T.~Vicsek.
\newblock Evolution of the social network of scientific collaborations.
\newblock {\em Physica A: Statistical mechanics and its applications},
  311(3):590--614, 2002.

\bibitem{barreno2006can}
M.~Barreno, B.~Nelson, R.~Sears, A.~D. Joseph, and J.~D. Tygar.
\newblock Can machine learning be secure?
\newblock In {\em Proceedings of the 2006 ACM Symposium on Information,
  computer and communications security}, pages 16--25. ACM, 2006.

\bibitem{boyd2004convex}
S.~Boyd and L.~Vandenberghe.
\newblock {\em Convex optimization}.
\newblock Cambridge university press, 2004.

\bibitem{bruckner2011stackelberg}
M.~Br{\"u}ckner and T.~Scheffer.
\newblock Stackelberg games for adversarial prediction problems.
\newblock In {\em Proceedings of the 17th ACM SIGKDD}, pages 547--555. ACM,
  2011.

\bibitem{budak2011limiting}
C.~Budak, D.~Agrawal, and A.~El~Abbadi.
\newblock Limiting the spread of misinformation in social networks.
\newblock In {\em Proceedings of the 20th international conference on World
  wide web}, pages 665--674. ACM, 2011.

\bibitem{burges1998tutorial}
C.~J. Burges.
\newblock A tutorial on support vector machines for pattern recognition.
\newblock {\em Data mining and knowledge discovery}, 2(2):121--167, 1998.

\bibitem{cheng2015antisocial}
J.~Cheng, C.~Danescu-Niculescu-Mizil, and J.~Leskovec.
\newblock Antisocial behavior in online discussion communities.
\newblock In {\em International Conference on Weblogs and Social Media}, pages
  61--70, 2015.

\bibitem{colson2007overview}
B.~Colson, P.~Marcotte, and G.~Savard.
\newblock An overview of bilevel optimization.
\newblock {\em Annals of operations research}, 153(1):235--256, 2007.

\bibitem{cormack2008email}
G.~V. Cormack et~al.
\newblock Email spam filtering: A systematic review.
\newblock {\em Foundations and Trends{\textregistered} in Information
  Retrieval}, 1(4):335--455, 2008.

\bibitem{dalvi2004adversarial}
N.~Dalvi, P.~Domingos, S.~Sanghai, D.~Verma, et~al.
\newblock Adversarial classification.
\newblock In {\em Proceedings of the tenth ACM SIGKDD}, pages 99--108. ACM,
  2004.

\bibitem{du2013scalable}
N.~Du, L.~Song, M.~G. Rodriguez, and H.~Zha.
\newblock Scalable influence estimation in continuous-time diffusion networks.
\newblock In {\em Advances in neural information processing systems}, pages
  3147--3155, 2013.

\bibitem{du2013uncover}
N.~Du, L.~Song, H.~Woo, and H.~Zha.
\newblock Uncover topic-sensitive information diffusion networks.
\newblock In {\em Artificial Intelligence and Statistics}, pages 229--237,
  2013.

\bibitem{du2012learning}
N.~Du, L.~Song, M.~Yuan, and A.~J. Smola.
\newblock Learning networks of heterogeneous influence.
\newblock In {\em Advances in Neural Information Processing Systems}, pages
  2780--2788, 2012.

\bibitem{gomez2012influence}
M.~Gomez-Rodriguez and B.~Sch{\"o}lkopf.
\newblock Influence maximization in continuous time diffusion networks.
\newblock In {\em Proceedings of the 29th International Coference on
  International Conference on Machine Learning}, pages 579--586. Omnipress,
  2012.

\bibitem{holcomb2013news}
J.~Holcomb, J.~Gottfried, and A.~Mitchell.
\newblock News use across social media platforms.
\newblock {\em Pew Research Journalism Project}, 2013.

\bibitem{kempe2003maximizing}
D.~Kempe, J.~Kleinberg, and {\'E}.~Tardos.
\newblock Maximizing the spread of influence through a social network.
\newblock In {\em Proceedings of the ninth ACM SIGKDD}, pages 137--146. ACM,
  2003.

\bibitem{Laszka15}
A.~Laszka, Y.~Vorobeychik, and X.~Koutsoukos.
\newblock Optimal personalized filtering against spear-phishing attacks.
\newblock In {\em AAAI Conference on Artificial Intelligence}, 2015.

\bibitem{li2014feature}
B.~Li and Y.~Vorobeychik.
\newblock Feature cross-substitution in adversarial classification.
\newblock In {\em Advances in neural information processing systems}, pages
  2087--2095, 2014.

\bibitem{li2016general}
B.~Li, Y.~Vorobeychik, and X.~Chen.
\newblock A general retraining framework for scalable adversarial
  classification.
\newblock {\em arXiv preprint arXiv:1604.02606}, 2016.

\bibitem{Lichman:2013}
M.~Lichman.
\newblock {UCI} machine learning repository, 2013.

\bibitem{lowd2005adversarial}
D.~Lowd and C.~Meek.
\newblock Adversarial learning.
\newblock In {\em Proceedings of the eleventh ACM SIGKDD}, pages 641--647. ACM,
  2005.

\bibitem{mei2015security}
S.~Mei and X.~Zhu.
\newblock The security of latent dirichlet allocation.
\newblock In {\em Artificial Intelligence and Statistics}, pages 681--689,
  2015.

\bibitem{mei2015using}
S.~Mei and X.~Zhu.
\newblock Using machine teaching to identify optimal training-set attacks on
  machine learners.
\newblock In {\em AAAI Conference on Artificial Intelligence}, pages
  2871--2877, 2015.

\bibitem{tsai2012security}
J.~Tsai, T.~H. Nguyen, and M.~Tambe.
\newblock Security games for controlling contagion.
\newblock In {\em AAAI Conference on Artificial Intelligence}, 2012.

\bibitem{ugander2011anatomy}
J.~Ugander, B.~Karrer, L.~Backstrom, and C.~Marlow.
\newblock The anatomy of the facebook social graph.
\newblock {\em arXiv preprint arXiv:1111.4503}, 2011.

\bibitem{vorobeychik2015securing}
Y.~Vorobeychik and J.~Letchford.
\newblock Securing interdependent assets.
\newblock {\em Autonomous Agents and Multi-Agent Systems}, 29(2):305--333,
  2015.

\bibitem{wallinga2004different}
J.~Wallinga and P.~Teunis.
\newblock Different epidemic curves for severe acute respiratory syndrome
  reveal similar impacts of control measures.
\newblock {\em American Journal of Epidemiology}, 160(6):509--516, 2004.

\bibitem{watts1998collective}
D.~J. Watts and S.~H. Strogatz.
\newblock Collective dynamics of small-world networks.
\newblock {\em Nature}, 393(6684):440--442, 1998.

\bibitem{zorichmathematical}
V.~A. Zorich and R.~Cooke.
\newblock Mathematical analysis i. 2004.

\end{thebibliography}
\end{document}